\documentclass[12pt]{article}
\usepackage{graphicx}
\usepackage{latexsym,amsfonts, amssymb,epsfig}
\usepackage{amssymb,amsmath,amsthm}
\usepackage[english]{babel}

\begin{document}

\begin{center}
{\Large\bf Compacton-like solutions to some nonlocal
hydrodynamic-type model}\footnote{The research was supported by the
AGH local grant}

\vspace{10mm}

{\it {\large\bf Vsevolod Vladimirov}\footnote{Tel.: (04812) 617 32
83; fax: (04812) 617 31 65} \footnote{{\it E-mail address:}
vsevolod.vladimirov@gmai.com }
\\
 \vspace{5mm}

Faculty of Applied Mathematics \\
  University of Science and Technology\\
Mickiewicz Avenue 30, 30-059 Krak\'{o}w, Poland \\
 [2ex] }

 \end{center}

\vspace{10mm}

{ \footnotesize {\bf Abstract.  }We study the appearance of
compacton-like solutions within the  hy\-dro\-dy\-na\-mic-type model
taking into account effects of spatial non-locality }

 \vspace{3mm}

 \noindent{\bf PACS codes:} 02.30.Jr; 47.50.Cd; 83.10.Gr

\vspace{3mm}

 \noindent {\bf Keywords:} Wave patterns; compactons  hydrodynamic-type
 model; spatial and temporal non-localities

\section{ Introduction }

 In this paper    evolutionary PDEs
 describing wave patterns with compact support are studied.
 Very often wave patterns play  key rules in  nonlinear transport
 phenomena \cite{Shkad,Davydov,spindeton2}.
One of the most advanced mathematical theory dealing with wave
patterns'  formation and evolution is
 the soliton theory. The well-known Korteveg
  -- de Vries  equation (KdV)
 \begin{equation}\label{KdV}
u_t+\beta\,u\,u_x+u_{xxx}=0,
 \end{equation}
possesses a one-parameter family of solutions describing
exponentially localized wave patterns  called {\it solitons}
\cite{Dodd}:
\begin{equation}\label{KdVsol}
u(t,\,\,x)=\frac{12\,a^2}{\beta}\,sech^2{\left[a(x-4\,a^2
\,t)\right]}.
\end{equation}
Solitons  demonstrate many outstanding features. Being the solutions
to non-linear evolutionary equation (\ref{KdV}), they manifest
"elastic properties" during the collisions. Besides, any
sufficiently smooth initial data possessing a finite energy norm
creates a chain of solitons moving with different velocities. All
this is recognized as a consequence of the complete integrability of
equation (\ref{KdV}) and existence of infinite hierarchy of
conservation laws.

 Actually the solitons have been associated with a number of
phenomena observed in science and technics. However, solution
(\ref{KdVsol}) cannot be fully identified with any kind of solitary
waves occurring in natural phenomena since the wave pattern
described by this  formula at fixed $t$ extends to $\pm \infty$.

In recent years there have been discovered another type of solitary
waves supported by the following generalization of the KdV hierarchy
{\cite{Ros_93}}:
\begin{equation}\label{Kmn}
K(m,\,n)=u_t+\left(u^m\right)_x+\left(u^n\right)_{xxx}=0,\quad
m,\,\,n\,\geq\,2.
\end{equation}
In case when $m=2,\,\,n=2$  a generalized solution to this equation
is described by the  following formula \cite{Ros_93}:
\begin{equation}\label{companalsol}
u(t,\,x)=\left\{\begin{array}{c}
\frac{4\,D}{3}\,\cos^2{\left[(x-D\,t)/4\right]} \,\,\,
\mbox{when}\,\,\, |x-D\,t|\leq{2\,\pi}, \\
0  \,\,\,\mbox{when}\,\,\, |x-D\,t|\geq\,{2\,\pi}.\end{array}
\right.
\end{equation}
Solutions of this sort are rather typical to the whole $K(m,\,n)$
hierarchy. All of them vanish outside some compact  domain and hence
they are referred to as {\it compactons}. Study of $K(m,\,n)$
hierarchy is actually in progress and there has already been
realized that most of these equations are not completely integrable
and do not possess an infinite set of conservation laws.
Nevertheless, compacton-supporting equations inherit many features
of equations belonging to the  KdV hierarchy. In particular, a
sufficiently smooth perturbations taken as Cauch\'{y} data give rise
to a chain of compactons \cite{Ros_93,Pik_Ros1,Pik_Ros2,Ros_2006}.
Besides, compactons manifest almost perfectly elastic features
during the mutual collisions  \cite{Ros_93,Pik_Ros2}.

It should be emphasized, yet, that  most papers dealing with the
subject in question, are concerned with the compactons being the
solutions to either completely integrable equations, or those which
produce a completely integrable ones when being reduced onto subset
of a traveling wave (TW) solutions \cite{Ros_93,OlvRosenau,Ros_96}.

In this paper  compacton-like solutions to  the hydrodynamic-type
model taking into account  the  effects of spatial non-locality are
considered. The model occurs to share the compacton-like solutions
with a Hamiltonian system but only accidentally, i.e. at certain
values of the parameters.  In spite of such restriction, existence
of this type of solutions seems to be significant for  their
appearance is connected with the presence of non-local effects and
rather cannot be manifested in any local hydrodynamic model.
Besides, localized invariant solutions sometimes manifest attractive
features and when this is the case, they can be treated as some
universal mechanism of the energy transfer in media with internal
structure leading to the given type of the hydrodynamic-type system.

The structure of the paper is following.  In  section~2 we introduce
dynamic equations of state (DES) taking into account  effects of
spatial and temporal non-localities.   In section~3 we perform the
qualitative analysis of  a hydrodynamic system of balance equations
closed by DES   taking into account  effects of spatial non-locality
and state the conditions leading to the existence of compacton-like
TW solutions. In the last section we summarize the results obtained,
discuss open problems and prospects of further developing the actual
study.

\section{Non-local hydrodynamic-type models}

We are going to analyze the existence of compacton-like solutions
within the hy\-d\-ro\-dy\-na\-mic-type models taking into account
non-local effects. These effects are manifested when an intens pulse
loading (impact, explosion, etc.) is applied to media possessing
internal structure on mesoscale. Description of the non-linear waves
propagation in such media depends in essential way on the ration of
a characteristic size $d$ of elements of the medium structure to the
characteristic length $\lambda$ of the wave pack. If $d/\lambda$  is
of the order of  $O(1)$ then the basic concepts of continuum
mechanics are not applicable and one should use the description
based e.g. on the element dynamics methods.  The models studied in
this paper apply when the ratio $d/\lambda$ is much less than unity
and therefore the continual approach is still valid, but it is not
as small that we can ignore the presence of internal structure.

As it is shown in \cite{VakhKul}, in the long wave approximation the
balance equations for  momentum and mass retain their classical
form, which in the one-dimensional case can be written as follows:
\begin{eqnarray}
u_t+p_x=0, \label{bal_imp} \\
\rho_t+\rho^2\,u_x=0. \label{bal_mass}
\end{eqnarray}
Here $u$ is mass velocity, $\rho$ is density, $p$ is pressure, $t$
is time, $x$ is mass (Lagrangean) coordinate.  So the whole
information about the presence of structure in this approximation is
contained in DES which should be added to system
(\ref{bal_imp})-(\ref{bal_mass}) in order to make it closed.

It is of common knowledge that a unified dynamic equation of state
well enough describing the behavior of condensed media in a wide
range of changes of pressure, density and regimes of load (unload)
actually does not exist. Various particular equations of state
describing dynamical behavior of  structural media are derived by
means of different techniques. There exist, for example, several
generally accepted DES in mechanics of heterogeneous media, derived
on pure mechanical ground (see e.g. \cite{Lyakhov,Nakoryakov}).
Contrary, in papers \cite{DeGroot,Lykov,Dan_93,Dan_Vl_96}
derivations of DES are based on the non-equilibrium thermodynamics
methods. Both, mechanical and thermodynamical approaches under the
resembling assumptions give rise to similar DES, stating the
functional dependencies between $p,\,\,\rho$ and their partial
derivatives.

There is also a number of works aimed at deriving the DES  on the
basis of the statistical  theory of irreversible processes (see
\cite{ZubarevTishch,Rudyak} and references therein).  It is rather
firmly stated within this approach that DES for complicated
condensed matter, being far from the state of thermodynamic
equilibrium, takes the form of integral equations, linking together
generalized thermodynamical fluxes $I_n$ and generalized
thermodynamical forces $L_m$, causing these fluxes:
\begin{equation}
I_n=f_n(L_k)+\int_{-\infty}^t\,d\,t'\int_{-\infty}^{+\infty}
K_{m\,n}\left(t,\,t';x,x' \right)\,g_m\left[L_k(t',\,x') \right].
\end{equation}
Here $K_{m\,n}\left(t,\,t';x,x' \right)$ is the  kernel of
non-locality, which can be calculated, in principle, by solving
dynamic problem of structure's elements interaction. Yet, such
calculations are extremely difficult and very seldom are seen
through to the end. Therefore we follow a common practice
\cite{Joseph,Peerlings} and use some model kernels describing well
enough the main properties of the non-local effects, in particular,
the fact that these effects vanish rapidly as $|t-t'|$ and $|x-x'|$
grow.

DES derived here are based on the  following relation between the
pressure and density:
\begin{equation}\label{gen_sptemp}
p(t,\,x)=f\,\left[\rho\left(t,\,x
\right)\right]+\int_{-\infty}^{t}\left\{\int_{-\infty}^{+\infty}
K\left(t-t',\,\,x-x'   \right)\,g\left[\rho\left(t',\,x'
\right)\right]\,d\,x'\right\}\,d\,t'.
\end{equation}
Let us first assume that effects of spatial non-locality are
unimportant. In this case the kernel of non-locality can be
presented as $K\left(t-t' \right)\,\delta\left(x-x'\right)$ and the
flux-force relation (\ref{gen_sptemp}) takes the following form:
\begin{equation}\label{gen_temp}
p(t,\,x)=f\,\left[\rho\left(t,\,x \right)\right]+\int_{-\infty}^{t}
K\left(t-t' \right)\,g\left[\rho\left(t',\,x \right)\right]\,d\,t'.
\end{equation}
Since we rather do not want to deal with the system of
integro-differential equations, our next step is to extract some
acceptable kernels enabling to pass from (\ref{gen_temp}) to pure
differential relations. Differentiating   equation (\ref{gen_temp})
with respect to the temporal variable we obtain:
\begin{equation}\label{temp_t}
p_t=\dot{f}\,\left[\rho\right]\,\rho_t+K(0)\,g\left[
\rho\right]+\int_{-\infty}^{t} K_t\left(t-t'
\right)\,g\left[\rho\left(t',\,x \right)\right]\,d\,t'.
\end{equation}
Equation(\ref{temp_t}) is equivalent to a pure differential one
provided that function $K(z)$ satisfies  equation
$\dot{K}(z)=c\,K(z)$.
%
In this case we can obtain from (\ref{gen_temp}), (\ref{temp_t})
the following differential equation:
\begin{equation}\label{onerelax}
\tau\left\{p_t-\dot{f}\,[\rho]\,\rho_t\right\}=\tau\,A\,g\left[\rho\right]+f\left[\rho\right]-p.
\end{equation}
This equation corresponds to fading memory kernel
$K(z)=A\,\exp\,[-\frac{z}{\tau}]$. For
$A=1,\,\,f\left[\rho\right]=\chi\rho^{n+1},\,\,g\left[\rho\right]=-\sigma\rho^{n+1}$,
we get the following DES:
\begin{equation}\label{onerelDES}
\tau\left\{p_t-\chi\,(n+1)\,\rho^n\,\rho_t\right\}=\kappa\,\rho^{n+1}-p.
\end{equation}
where $\kappa=\chi-\sigma\,\tau$. Equations coinciding with
(\ref{onerelDES}) under certain additional assumptions are widely
used to  describe nonlinear processes in multi-component media with
one relaxing process in the elements of structure
\cite{Lyakhov,Dan_93}.

In more complicated cases,  e.g. when more than one relaxing process
should be taken into account, another kernels are to be applied. Let
us look for the kernels leading from (\ref{gen_temp}) to pure
differential equation under the assumption that such passage is
possible with the help of two differentiations. Taking derivative of
(\ref{temp_t}) with respect to the temporal variable we get:
\begin{equation}\label{temp_tt}
p_{tt}=\left[\dot{f}\,\left[\rho\right]\,\rho_t\right]_t+K(0)\,\dot{g}\left[
\rho\right]\,\rho_t+\dot{K}[0]\,g\left[\rho\right]+\int_{-\infty}^{t}
K_{tt}\left(t-t' \right)\,g\left[\rho\left(t',\,x
\right)\right]\,d\,t'.
\end{equation}
So now we are interested in whether the linear combination
$h\,p_{tt}+s\,p_{t}+p$ can be expressed as a pure differential
equation. It is easy to see that the integral term corresponding to
this combination is as follows:
\[
\int_{-\infty}^{t}\left\{ h\,K_{tt}\left(t-t'
\right)+s\,K_{t}\left(t-t' \right)+K\left(t-t'
\right)\right\}\,g\left[\rho\left(t',\,x \right)\right]\,d\,t'.
\]
Our requirement will be addressed if $K[z]$ satisfies the equation
\begin{equation}\label{K_zz}
h\,K_{zz}\left(z \right)+s\,K_{z}\left(z\right)+K\left(z \right)=0.
\end{equation}
If this is so, then  DES takes on the form
\begin{eqnarray}\label{soge_temp}
h\,p_{tt}+s\,p_{t}+p=f(\rho)+\left[s\,K(0)+h\,K_t(0)\right]\,L(\rho)+
\\
   +\left[s\,\dot{f}(\rho)+h\,K(0)\dot{g}(\rho)\right]\rho_t
   +h \left[\dot{f}(\rho)\rho_{tt}+\ddot{f}(\rho)\rho_{t}^2\right].
   \nonumber
\end{eqnarray}

Their are two cases for which requirement of fading memory holds
true:
\[
K[z]=\alpha\,e^{-\frac{z}{\tau_1}}+\beta\,e^{-\frac{z}{\tau_2}},\quad
h=\tau_1\,\tau_2,\quad s=\tau_1\,+\,\tau_2,
\]
and
\[
K[z]=e^{-\frac{z}{\tau_1}}\,\sin\left[\frac{z}{\tau_2}+\gamma\right],\quad
h=\frac{\tau_1^2\,\tau_2^2}{\tau_1^2\,+\,\tau_2^2},\quad
s=\frac{2\,\tau_1\,\tau_2^2}{\tau_1^2\,+\,\tau_2^2}.
\]
In the first case we have the kernel describing media with two
relaxing processes, while in the second case the  kernel describes
relaxing media with oscillating component in the elements of
structure. Note that in  both  cases we  get formally identical DES
(\ref{soge_temp}), yet the parameters corresponding to them satisfy
quite  different inequalities. In the first case the parameters
$h,\,\,s$ obey the inequalities $s^2>4\,h>0$, while in the second
case the inequalities are as follows: $0<s^2<4\,h$. So the
parameters lie in the distinct sets of parameter space and this fact
occurs to be crucial since the behavior of the system of balance
equation (\ref{bal_imp})-(\ref{bal_mass}) closed by  DES
(\ref{soge_temp}) is extremely sensible on the values of these
parameters. \cite{Dan_Vl_96,Sym_97}.

Now let us address the case of pure spatial non-locality. Following
\cite{Peerlings}, we shall use the kernel of the  form
$K\left[t-t',\,x-x'
\right]=\hat\sigma\,\exp{\left[-\left(\frac{x-x'}{l}\right)^2\right]}\cdot\delta\left[t-t'
\right]$ giving the equation
\begin{equation}\label{Int_spt}
p=f\,\left(\rho
\right)+\hat\sigma\,\int_{-\infty}^{+\infty}\,e^{-\left[\frac{x-x'}{\l}\right]^2}
\,g\left[\rho\left(t,\,x' \right)\right]\,d\,x'.
\end{equation}
Since the function
$\exp{\left[-\left(\frac{x-x'}{l}\right)^2\right]}$ extremely
quickly approaches zero as $|x-x'|$ grows, we  use the following
approximation for function $g\left[{\rho}\left(t,\,x'
\right)\right]$ inside the inner integral:
\begin{equation}\label{grad}
g\left[{\rho}\left(t,\,x' \right)\right]=g\left[{\rho}\left(t,\,x
\right)\right]+\bigl\{g\left[{\rho}\left(t,\,x
\right)\right]\bigr\}_x\,\frac{(x'- x)}{1\,!}+
\bigl\{g\left[{\rho}\left(t,\,x
\right)\right]\bigr\}_{xx}\,\frac{(x'-x)^2}{2\,!}+ O(|x'-x|^3).
\end{equation}
Dropping out the term $O(|x'-x|^3)$ and integrating over $d\,x'$, we
get
\begin{equation}\label{spat_perlings}
p= f\left[{\rho}(t,\,x)\right]+\sigma_0\,g\left[{\rho}\left(t,\,x
\right)\right]+\sigma_2\,\left\{g\left[{\rho}\left(t,\,x
\right)\right]  \right\}_{xx}
\end{equation}
where
\[
\sigma_0=\hat\sigma\,\int_{-\infty}^{+\infty}\,e^{-\left[\frac{x-x'}{\l}\right]^2}\,d\,x',
\quad
\sigma_2=\hat\sigma\,\int_{-\infty}^{+\infty}\,\frac{(x'-x)^2}{2\,!}\,
e^{-\left[\frac{x-x'}{\l}\right]^2}\,d\,x'.
\]

It is obvious, that one can easily combine two types of
non-localities by taking the kernel in the form of the product
\[
\bar{K}\left[t-t',\,\,x-x'
\right]=K\left[t-t'\right]\cdot\exp{\left[-\left(\frac{x-x'}{l}\right)^2\right]},
\]
with $K\left[t-t'\right]$ satisfying (\ref{K_zz}). Employment of
technics used in the last two items enables to obtain the DES taking
into account both effects of spatial and temporal non-localities:
\begin{eqnarray}\label{soge_spatemp}
h\,p_{tt}+s\,p_{t}+p=f[\rho]+\left[s\,K[0]
+h\,K_t[0]\right]\,\left[\sigma_0\,g[\rho]+\sigma_2\left(\dot{g}[\rho]\,\rho_x
 \right)_x\right]+ \\
    \qquad\qquad\qquad\qquad+s\,\dot{f}[\rho]\,\rho_t+
    h\,K[0]\left[\sigma_0\,g[\rho]
    +\sigma_2\left(\dot{g}[\rho]\,\rho_x
 \right)_x\right]_t
   +h \left[\dot{f}[\rho]\rho_t\right]_t. \nonumber
\end{eqnarray}

Let us note that DES with higher derivatives were employed in many
papers dealing with the structured media. Qualitative and numerical
investigations undertaken in \cite{Sym_97,ROMP_1999,ROMP_2000} show
that the modeling system taking into account both types of
non-localities possesses a rich set of the TW solutions, including
periodic, quasi-periodic, multi-periodic, soliton-like and chaotic
regimes.

\section{Compacton-like solutions within the hyd\-ro\-dy\-na\-mic-type
model taking into account spatial non-localitiy}

Before we start to  "capture" compacton-like solutions within the
hydrodynamic model,  it is desired to get a clear idea on what is a
compactons from geometric point of view. In paper \cite{ROMP_08}
such analysis has been performed for the equations belonging to
Rosenau-Hyman $K(n,\,m)$ hierarchy and this is the way things stand.
 Being treated as function of a single value $\xi-x-D\,t$, solution (\ref{companalsol})
 satisfies some second-order ODE which is obtained from
 (\ref{Kmn}) by means of the ansatz $u(t,\,x)=U(\xi).$  Substituting
 this ansatz into the source equation one gets, after one
 integration and some manipulation, a second-order Hamiltonian system
  \cite{ROMP_08}. This system occurs to have a one-parameter family
  of periodic  phase trajectories and the homoclinic trajectory surrounding
  them. The last one just corresponds to compacton.

  In  the case of KdV the similar procedure leads to the Hamiltoinan
  system with the same geometry of the phase trajectories.
  The difference between two phase portraits arises from the fact that in
  the latter case the homoclinic loop  is the bi-asymptotic
  trajectory of a simple saddle while the trajectory representing
  the compacton is the separatrix of a saddle settled on the
  line of singular points of the  vector field.
  Therefor the homoclinic trajectlry is penetrated in a finite
  "time" and the compacton solution (\ref{companalsol}) consists of
  non-zero part with compact support, glued with zero solution
  corresponding to the saddle point.

Let us note at the end of this short introduction, that we do not
distinguish solutions having a compact support and those which can
be made so by proper change of variables.

Now let us consider the system (\ref{bal_imp}), (\ref{bal_mass})
closed by DES
\begin{equation}\label{DEqS}
p=f[\bar{\rho}]+\kappa\,\bar{\rho}^{n+2}+\sigma\,\left[\bar{\rho}^{n+1}\bar{\rho}_x
\right]_x
\end{equation}
corresponding to pure spatial non-locality. Here
$\bar{\rho}=\rho-\rho_0,$ where $0<\rho_0$ is a constant,
$f[\bar{\rho}]$ is a  function which will be defined later on.  We
use the  ansatz
\begin{equation}\label{TW_ans} u(t,x)=U(x-D\,t)\equiv
U(\xi),  \qquad \bar{\rho}(t,\,x)=R(\xi),
\end{equation}
enabling to factorize system (\ref{bal_mass}),
 (\ref{bal_imp}), closed by the DES (\ref{DEqS}).
Inserting (\ref{TW_ans}) into equation (\ref{bal_mass}), we get the
following quadrature:
\begin{equation}\label{bal_mas_q}
U=\frac{D}{\rho_0}-\frac{D}{R[\xi]+\rho_0}.
\end{equation}
 Constant of integration have been chosen in such a way that
$u(t,\,\pm\,\infty)=0.$

Using the ansatz (\ref{TW_ans}), we express DES in new invariant
variables. Inserting (\ref{DEqS})   into (\ref{bal_imp}), using the
formula (\ref{bal_mas_q}) and integrating once the expression
obtained this way we pass, after some manipulation, to the following
second order ODE
\begin{equation}\label{SODE}
\frac{D^2}{R+\rho_0}+f[R]+\kappa\,R^{n+2}+\sigma\,\left[R^{n+1}\,R'\right]'=E=\frac{D^2}{\rho_0}+f[0].
\end{equation}
It is obvious that above equation can be re-written as an equivalent
dynamical system. To do this, we define a new function $W=-R'$. Next
we introduce new independent variable $T$ such that
$\frac{d}{d\,T}=\sigma\,R^{n+1}\,\varphi[R]\frac{d}{d\,\omega},$
where $\varphi[R]$ is an integrating factor which is incorporated in
order to make the system Hamiltonian. With this notation we get the
following dynamical system equivalent to (\ref{SODE}):
\begin{equation}\left\{\begin{array}{l}
\frac{d\,R}{d\,T}=-\sigma\,\varphi[R]\,R^{n+1}\,W=-\frac{\partial\,H[R,W]}{\partial\,W}, \label{hmlt_DS} \\
\frac{d\,R}{d\,T}=\varphi[R]\left[
\sigma(n+1)R^n\,W^2+f(R)-f(0)+\kappa
R^{n+2}-\frac{D^2R}{\rho_0\left(R+\rho_0\right)}
\right]=\frac{\partial\,H[R,W]}{\partial\,R}.
\end{array} \right.
\end{equation}
Solving the first equation of system (\ref{hmlt_DS}) with respect to
function $H[R,\,W]$ we obtain:
\begin{equation}\label{hmlt_aux}
H[R,\,W]=\sigma\,\varphi[R]\,R^{n+1}\frac{W^2}{2}+\theta[R].
\end{equation}
Now, comparing the RHS of second equation of system (\ref{hmlt_DS})
with partial derivative of (\ref{hmlt_aux}) with respect to $R$, we
get the system
\begin{eqnarray*}
R\,\varphi^{\prime}[R]=(n+1)\varphi[R], \\
\theta^{\prime}[R]=\varphi[R]\,\left\{f(R)-f(0)+\kappa
R^{n+2}-\frac{D^2R}{\rho_0\left(R+\rho_0\right)}\right\}.
\end{eqnarray*}
The first equation is satisfied by the function
$\varphi[R]=C\,R^{n+1}$. For convenience we put $C=2$. Inserting
$\varphi[R]$ in the second equation we obtain the quadrature
\[
\theta(R)=2 \int\,R^{n+1}\left\{f(R)-f(0)+\kappa
R^{n+2}-\frac{D^2R}{\rho_0\left(R+\rho_0\right)}\right\}\,d\,R.
\]
So the Hamiltonian function is finally expressed as  follows:
 \begin{equation}\label{hmlt_fct}
H[R,\,W]=\sigma\,R^{2(n+1)}{W^2}+2
\int\,R^{n+1}\left\{f(R)-f(0)+\kappa
R^{n+2}-\frac{D^2R}{\rho_0\left(R+\rho_0\right)}\right\}\,d\,R.
\end{equation}

The Hamiltonian function (\ref{hmlt_fct}) is a first integral of the
system (\ref{hmlt_DS}), so every phase trajectory can be presented
in the form
\begin{equation}\label{traject_1}
W^2=\frac{1}{\sigma\,R^{2(n+1)}}\,\left\{K-2
\int\,R^{n+1}\left[f(R)-f(0)+\kappa
R^{n+2}-\frac{D^2R}{\rho_0\left(R+\rho_0\right)}\right]\,d\,R
\right\},
\end{equation}
where $K$ is the  constant value of the Hamiltonian on a particular
trajectory. Now let us analyze formula (\ref{traject_1}). If we want
to have a closed trajectory approaching the origin, then we must
properly choose the constant $K$ and ''suppress'' all the singular
terms by the proper choice of function $f(R)$. It can be easily
shown by induction that the following decomposition for the last
term inside the integral takes place:
\[
\frac{R^{n+2}}{R+\rho_0}=R^{n+1}-\rho_0\,R^{n}+...+(-1)^k\,\rho_0^k\,R^{n+1-k}+...
+(-1)^{n+1}\,\rho_0^{n+1}+(-1)^{n+2}\frac{\rho_0^{n+2}}{R+\rho_0}.
\]
Hence
\[
W^2=\frac{1}{\sigma\,R^{2(n+1)}}\,\left\{K-2
\int\,R^{n+1}\left[f(R)-f(0)+\kappa R^{n+2}\right]\,d\,R- \right.
\]
\[
\left.
-\frac{2\,D^2}{\rho_0}\left(\frac{R^{n+2}}{n+2}-\rho_0\frac{R^{n+1}}{n+1}+...
+(-\rho_0)^{n+1}\,R+(-\rho_0)^{n+2}\log\left(R+\rho_0 \right)\right)
\right\}.
\]
From this we conclude that the last term in (\ref{traject_1}) always
produces singularities. More precisely,  singularities are connected
with monomials $R^m$ when $m<2(n+1)$ and with the logarithmic term.
Therefore the last term in (\ref{traject_1}) should be rather
removed by  the proper choice of $f(R)$.

A simple  analysis shows that function
\[
f(R)=f_0+\frac{A\,R}{\rho_0\,\left(R+\rho_0\right)}+g_1\,R^{n+1}+g_2\,R^{n+2}
\]
with $A>0$ will suppress singularity provided that $D=\pm\sqrt{A}$.
In fact, in this case
\begin{equation}\label{traject_2}
W^2=\frac{1}{\sigma\,R^{2(n+1)}}\,\left\{K-R^{2(n+1)}\left[\frac{2
g_1\,R}{2n+3}+\frac{\bar{g}_2\,R^2}{n+2} \right] \right\},
\end{equation}
where $\bar{g}_2=g_2+\kappa$. With such a choice the only trajectory
approaching  the origin corresponds to $K=0.$ If in addition
$g_1=-\alpha_1<0$ and $\bar{g_2}=\alpha_2>0,$ i.e.
\begin{equation}\label{comp_DES}
f(R)=f_0+\frac{A\,R}{\rho_0\,\left(R+\rho_0\right)}-\alpha_1\,R^{n+1}+\alpha_2\,R^{n+2}
\end{equation}
then
\begin{equation}\label{traject_3}
W=\pm\,\frac{1}{\sqrt{\sigma}}\sqrt{\frac{2\,\alpha_1}{2n+3}\,R-\frac{\alpha_2}{n+2}R^2}
\end{equation}
and there is the point $R_*=\frac{2(n+2)\alpha_1}{(2n+3)\alpha_2}$
in which the trajectory intersects the horizontal axis.

In fact, under the above assumptions we get the geometry which is
similar to that obtained when the member of $K(n,m)$ hierarchy is
reduced to an ODE describing the set of TW solutions \cite{ROMP_08}.
To show that, let us consider the system arising from (\ref{SODE})
under the above assumptions:
\begin{eqnarray}\label{CompDS}
\frac{d\,R}{d\,T}=-2\,\sigma\,W\,R^{2(n+1)}\qquad\qquad\qquad\qquad \label{comp_2} \\
\frac{d\,W}{d\,T}=2\,(n+1)\,\sigma\,W^2\,R^{2n+1}+2\,\alpha_2\,R^{2n+3}-2\,\alpha_1\,R^{2n+2}.
\nonumber
\end{eqnarray}
System (\ref{CompDS}) possesses two stationary points lying on the
horizontal axis:  the point $(0,\,0)$, and the point $(R_1,\,0)$,
where $R_1=\alpha_1/\alpha_2$. Analysis of the  Jacobi matrix shows
that $(R_1,\,0)$ is a center.  Moreover, $R_1<R_*$ when $n>-2$ so
the closed trajectory corresponding to $K=0$  encircles the domain
filled with the periodic trajectories.  Using the  asymptotic
decomposition of the  solution  represented by the homoclinic
trajectory near the origin, it can be shown that the  trajectory
reaches the stationary point $(0,\,0)$ in finite "time" so it does
correspond to the compacton.

Besides, we can get more direct evidence of the existence of
compacton-like solutions by integrating equation (\ref{traject_3}).
In fact, taking in mind that $W=-d\,R/d\,\xi$, we obtain the
following equation:
\[
\frac{1}{\gamma}\,\frac{d\,R}{\sqrt{1-\left(\frac{R}{\gamma}-1\right)^2}}=\delta\,d\,\xi,\]
where $\gamma=\frac{\alpha_1(n+2)}{\alpha_2(2n+3)},$
$\delta=\sqrt{\frac{\alpha_2}{\sigma(n+2)}}.$ Solving this equation
we get:
\begin{equation}\label{comp_spnl}
R[\xi]=\left\{
\begin{array}{l}
\gamma\,\left[1+\sin{\left(\delta\,(\xi-\xi_0)\right)}\right] \quad
\mbox{when} -\pi\leq\,2\,\delta\,(\xi-\xi_0)\leq\,3\,\pi
\\
0 \quad \quad \mbox{otherwise}.
\end{array}
\right.
\end{equation}

Using the formulae (\ref{DEqS})--(\ref{bal_mas_q}) and
(\ref{comp_spnl}), one can easily reproduce the functions
$u,\,\,\rho$ and $p$.

\section{Discussion}

In this work we have shown that hyd\-ro\-dy\-na\-mic system of
balance equations (\ref{bal_imp})-(\ref{bal_mass}) closed by DES
(\ref{DEqS}) possesses the compacton-like solutions. In contrast to
analogous solutions to most of the compacton-supporting equations,
the presented solutions do not form a one-parameter family. More
precisely, for fixed values of the parameters
 $A,\,\,\alpha_1$ and $\alpha_2$ in (\ref{comp_DES}) there exists exactly one pair of
 compactons moving with  velocity $\sqrt{A}$ in the opposite
 directions. Whether these solutions are of interest from the point
 of view of applications or not, depends on their stability and behavior
 during the mutual collisions. Discussion of these topics goes  beyond the
 scope of this paper. Let us mention, however, that the mere fact of  existence of compactons is the
 consequence of the non-local effects incorporation. To our best
 knowledge, this type of solutions does not exist in any local
 hydrodynamic-type model, i.e. the system of balance equations
 (\ref{bal_imp})-(\ref{bal_mass}) closed
 by the functional state equation $p=\Phi(\rho)$.
Let us  note that invariant TW solutions very often play rule of
intermediate asymptotics \cite{Barenblatt,Bedlewo}, attracting
near-by, not necessarily invariant, solutions. This feature
demonstrate compacton-like solutions of another non-local model
obtained when the system of balance equations
(\ref{bal_imp})-(\ref{bal_mass}) is closed by the DES
(\ref{onerelDES}), describing relaxing media \cite{ROMP_08}.
Solutions with compact supports appear in this model merely in
presence of the mass force. Exactly one compacton-like solution
occurs to  exist  for the given set of the parameters, yet  this
solution serves as an attractor for the wave packs created by the
wide class of the initial value problems \cite{ROMP_08}. In contrast
to the above mentioned relaxing model, incorporation of the effects
of spatial non-locality leads to the existence of a pair of
compacton-like solutions in absence of an external force. In fact,
it is shown for the first time  that a non-local hydrodynamic-type
model possesses more than one compacton-like solution and from now
on there exists the opportunity to investigate their behavior during
the  collisions.

There are some evidences in favor of the stable behavior of these
compactons and their elastic  collisions. The first is connected
with the fact that by proper choice of parameters $A,\,\,\alpha_1$
and $\alpha_2$  the evolutionarity conditions \cite{Landau}
$\partial\,p/\partial\rho>0,$ $\partial^2\,p/\partial^2\rho>0$ can
be fulfilled for DES (\ref{DEqS}). Besides, the  system  of balance
equations (\ref{bal_imp})-(\ref{bal_mass}) closed by the DES
(\ref{DEqS}) possesses at least two conservation laws. Though
actually it is not known how many conserved quantities assure the
stability of the wave patterns during the collisions, it is almost
certain that this number should not be infinite. For example, the
members of the Rosenau-Hyman $K(n,\,m)$ hierarchy, possessing with
certain four conserved quantities, collide almost elastically and so
is with the compacton-compacton and kovaton-compacton interactions
in the Pikovsky-Rosenau model \cite{Pik_Ros2}.

Another open question connected with the non-local hydrodynamic-type
models is as follows. It is unknown as yet whether the
compacton-like solutions exist within the hydrodynamic-type models
closed by the DES (\ref{soge_spatemp}) taking into account both the
effects of spatial and temporal non-localities. Our preliminary
analysis shows that incorporation of   terms connected with temporal
non-locality makes the model dissipative. So the Hamiltonian
formalism is of no use in this case and  investigations  of the
compacton-like regimes appearance should be based on some other
technics. Study of these questions is actually in progress.


\end{document}